\documentclass[10pt,prd,twocolumn,preprint,nofootinbib]{revtex4}

\usepackage{epsfig}
\usepackage{amsmath,amsfonts,amssymb}
\usepackage{t1enc}
\usepackage{verbatim}
\usepackage{float}
\usepackage{morefloats}
\usepackage{color}

\providecommand{\openone}{\leavevmode\hbox{\small1\kern-3.8pt\normalsize1}}

\usepackage{booktabs}
\usepackage[Q=yes,pverb-linebreak=no]{examplep}

\newcommand{\Vl}{V_L}
\newcommand{\Vr}{V_R}
\newcommand{\gl}{g_L}
\newcommand{\gr}{g_R}

\newcommand{\be}{\begin{equation}}
\newcommand{\ee}{\end{equation}}

\begin{document}

\title{\boldmath Top Quark Anomalous Couplings at the High-Luminosity Phase of the LHC \unboldmath}

\author{
Fr\'ed\'eric~D\'eliot$^{1}$,
Miguel C. N. Fiolhais$^{2,3}$,
Ant\'onio Onofre$^{4}$
\\[3mm]
{\footnotesize {\it 
$^1$ Institut de Recherche sur les lois Fondamentale de l'Univers, D\'epartement de Physique des Particules, CEA, Universit\'e Paris-Saclay, F-91191 Gif-sur-Yvette Cedex, France \\
$^2$ Science Department, Borough of Manhattan Community College, City University of New York, \\ 199 Chambers St, New York, NY 10007, USA \\
$^3$ LIP, Departamento de F\'{\i}sica, Universidade de Coimbra, 3004-516 Coimbra, Portugal \\
$^4$ Departamento de F\'{\i}sica, Universidade do Minho, 4710-057 Braga, Portugal \\
}}
}

\begin{abstract}
The combination of the latest and most precise measurements of several top quark properties is presented in this paper in order to establish allowed regions on anomalous contributions to the Lorentz structure of the $Wtb$ vertex. These measurements include single top production cross sections, $W$ boson helicity fractions and forward-backward asymmetries, both at Tevatron and at the Large Hadron Collider, up to a center-of-mass energy of 13~TeV. The results obtained at 95\% Confidence Level for the top quark anomalous couplings are compared with the limits extracted from a combination that includes the expected measurements at the future High-Luminosity run of the Large Hadron Collider.
\end{abstract}

\maketitle

\section{Introduction}
\label{sec:intro}

Since the beginning of operations, the Large Hadron Collider (LHC) has not found yet any solid sign of new physics beyond the Standard Model (SM) predictions. Even though all the experimental results from the LHC have shown remarkable consistency with those predicted by the SM, several questions remain unanswered. For example, it fails to explain the matter/anti-matter asymmetry of the Universe, the mass of neutrinos or to provide a candidate for dark matter. However, the fact that beyond the SM (BSM) phenomena may only exist at a higher energy scale, above the current and future reach of the LHC, imposes a strong limitation on the direct searches for new physics, such as a new fundamental particle or interaction. As an alternative, indirect searches offer a promising opportunity to grasp an early detection of a possible BSM physics signal. To this effect, both ATLAS and CMS have conducted extensive research programs over the past years to precisely measure observables that are highly sensitive to anomalous contributions to the $Wtb$ vertex. These anomalous couplings can be parameterized within an effective field theory approach, which includes dimension-six gauge-invariant effective operators~\cite{Buchmuller:1985jz,AguilarSaavedra:2008zc,AguilarSaavedra:2009mx}. In this framework, the most general $Wtb$ vertex is expressed as,
\begin{eqnarray}
\mathcal{L}_{Wtb} & = & - \frac{g}{\sqrt 2} \bar b \, \gamma^{\mu} \left( \Vl
P_L + \Vr P_R
\right) t\; W_\mu^- \nonumber \\
& - &  \frac{g}{\sqrt 2} \bar b \, \frac{i \sigma^{\mu \nu} q_\nu}{M_W}
\left( \gl P_L + \gr P_R \right) t\; W_\mu^- + \mathrm{h.c.} 
\label{ec:lagr}
\end{eqnarray}
The coupling $\Vl = V_{tb} \simeq 1$ and the anomalous couplings, $\Vr$, $\gl$, $\gr$ are dimensionless, complex, and equal to zero at tree level, in the SM. Despite the fact that anomalous couplings are absent in the SM at tree level, they may receive non-zero contributions from BSM physics effects. These anomalous contributions can be tested in top quark decays by measuring the $W$ boson helicity fractions and related observables, such as forward-backward asymmetries~\cite{AguilarSaavedra:2010nx,Aguilar-Saavedra:2015yza}. Single top quark production cross section measurements are also sensitive to anomalous contributions to the $Wtb$ vertex, and can be used in combination with the aforementioned observables to establish allowed regions on the anomalous couplings~\cite{Boos:1999dd,Najafabadi:2008pb,AguilarSaavedra:2008gt,Chen:2005vr,Zhang:2010dr}. It should also be mentioned that an effort was recently undertaken to establish common standards for the interpretation of top quark measurements within the SM effective field theory~\cite{AguilarSaavedra:2018nen}.

The High-Luminosity phase of the Large Hadron Collider (HL-LHC) is scheduled to start after 2025 and is projected to operate at a peak luminosity of \mbox{$\sim 5\times 10^{34}$~cm$^{-2}$m$^{-1}$}, roughly 2 to 2.5 times higher than the current instantaneous luminosity at the LHC. The expected significant increase in statistics at the HL-LHC may allow the detection of rare SM physics processes or possible new physics signals, such as the production of four top quarks, top-anti-top quark ($t\bar{t}$) resonances, Flavor Changing Neutral Currents (FCNC) top quark decays, Vector-Boson Fusion (VBF) Higgs production or the Higgs boson trilinear self-coupling, among others. The HL-LHC is also expected to have a significant impact on the increase of precision of several top quark properties measurements that have been under study at the LHC, such as the previously mentioned $Wtb$ observables. An extrapolation exercise is performed in this paper in order to estimate the expected increase of precision on the statistical and systematic uncertainties of these top quark observables. This exercise is performed assuming the SM hypothesis at a center-of-mass energy of 14~TeV and a total integrated luminosity of $3000$~fb$^{-1}$ at the HL-LHC.

In this paper, the most precise top quark measurements are used in a global fit to determine the anomalous couplings allowed regions at 95\% Confidence Level (CL). The allowed regions are presented in different scenarios of measurements combinations performing a global fit, in order to show the importance of each particular observable in narrowing the limits of the anomalous couplings. These results are directly compared with the limits extracted from the combination of the current most precise measurements and with the expected measurements obtained from the \mbox{HL-LHC} extrapolation exercise. 

\section{Current Measurements and High-Luminosity Extrapolation}
\label{sec:measurements}

The list of the $W$ boson helicity measurements used in the global fit to the $Wtb$ anomalous couplings is presented in Table~\ref{tab:whelicities}, both at Tevatron~\cite{Aaltonen:2012rz} and at the LHC~\cite{Aaboud:2016hsq}. The overall correlation coefficient between the longitudinal and left-handed (right-handed) $W$ boson helicity fraction measurements was found to be $\rho=-0.55$ ($\rho=-0.86$) at the LHC (Tevatron). These results are compatible with the SM prediction at next-to-next-to-leading order (NNLO) in QCD~\cite{Czarnecki:2010gb}. The helicity fractions $F_L$ and $F_0$ are more sensitive to $g_R$ than to $g_L$ and $V_R$, due to an interference term $V_L g_R^*$, which is not suppressed by the bottom quark mass, as happens for the $g_L$ and $V_R$ couplings~\cite{AguilarSaavedra:2006fy}. This linear term dominates over the quadratic one and makes $F_L$ and $F_0$ (and related quantities) very sensitive to $g_R$.

\begin{table}[h]
\begin{center}
\begin{tabular}{|c|c|c|}
\hline
   $W$ boson           &                  &                           \\ 
 helicity fractions    &   Tevatron       &  LHC ($\sqrt{s}=8$~TeV)   \\ 
\hline
   $F_0$               & $0.722\pm 0.081$ &  $0.709\pm0.019$   \\
   $F_L$               &      N/A         &  $0.299\pm0.015$   \\
   $F_R$               & $-0.033\pm0.046$ &         N/A        \\
\hline
\end{tabular}
\caption{List of measurements of $W$ boson helicities at Tevatron and at run 1 of the LHC.}
\label{tab:whelicities}
\end{center}
\end{table}

The measurements of the normal and transverse polarization forward-backward asymmetries,  $A_{FB}^{N}$ and $A_{FB}^{T}$, were also taken into account in the global fit together with the $A_{FB}^{\ell}$ asymmetry, associated with the  angle between the lepton momentum in the top quark rest frame and the top quark spin direction ($\theta_{\ell}$). These forward-backward asymmetries were measured by the ATLAS experiment~\cite{Aaboud:2017aqp}, using proton-proton collision data collected at a center-of-mass energy of 8~TeV, and are particularly sensitive to the anomalous couplings. In particular the imaginary part of $g_R$ has a linear dependence with the normal asymmetry and the top quark polarization ($P_t$), \emph{i.e.}
\begin{equation}
A_{FB}^{N}=0.64\times P_t \times Im(g_R)
\, , 
\end{equation}
assuming  $V_R=g_L=0$ and $V_L=1$~\cite{AguilarSaavedra:2010nx}.

\begin{table}[h]
\begin{center}
\begin{tabular}{|c|c|c|c|}
\hline
   Forward-backward           &     &    &         \\ 
   asymmetries  &  $A_{FB}^{N}$   & $A_{FB}^{T}$ &  $A_{FB}^{\ell}$  \\ 
\hline
  LHC ($\sqrt{s}=8$~TeV)      	& $-0.04\pm0.04$  &  $0.39\pm0.09$  &  $0.49\pm0.06$ \\
\hline
\end{tabular}
\caption{List of measurements of forward-backward asymmetries at the LHC.}
\label{tab:asymmetries}
\end{center}
\end{table}

Measurements of single top quark production cross sections were also considered in the combination. The list of experimental measurements on the $t$-, $Wt$-, and $s$-channel is presented in Table~\ref{tab:singletop} for Tevatron~\cite{Aaltonen:2015cra,CDF:2014uma} and for the LHC at different center-of-mass energies~\cite{Sirunyan:2016cdg,Sirunyan:2018lcp,Aaboud:2017pdi,Khachatryan:2016-023,Aad:2015eto,Aad:2015upn,Chatrchyan:2012ep,Chatrchyan:2012zca,Khachatryan:2016ewo}. These results are consistent with the SM predictions~\cite{Aliev:2010zk,Kant:2014oha,Brucherseifer:2014ama,Kidonakis:2015nna,Kidonakis:2011wy,Kidonakis:2010ux,Kidonakis:2010tc,Kidonakis:2012rm}. The measurement of the $t$-channel cross-section is especially important due to its sensitivity to the \mbox{Cabibbo-Kobayashi-Maskawa} quark mixing matrix element $V_{tb}$. This measurement provides the most direct and precise probe of this coupling. The $Wt$ associated production cross section is also relevant, given its size at the LHC, which helps to constrain new physics that may change the structure of the $Wtb$ vertex. Its measurements were also included in the global fit, as well as the different results for the $s$-channel single top quark production cross sections.

\begin{table}[h]
\begin{center}
\begin{tabular}{|c|c|c|c|}
\hline
   Single top           &     &    &         \\ 
   cross sections  &   $\sigma_t$ / pb  &   $\sigma_{Wt}$ / pb  &    $\sigma_{s}$ / pb   \\ 
\hline
  Tevatron          	& $2.25^{+0.29}_{-0.31}$   &  N/A   &$1.29^{+0.26}_{-0.24}$ \\
  LHC ($\sqrt{s}=7$~TeV)      	& $67.2\pm6.1$   & $16^{+5}_{-4}$  & $7.1\pm8.1$   \\
  LHC ($\sqrt{s}=8$~TeV)      	& $89.6^{+7.1}_{-6.3}$  & 23.1$\pm$3.6  & $4.8^{+1.8}_{-1.6}$   \\ 
  LHC ($\sqrt{s}=13$~TeV)      	& 238$\pm$32   & $63.1\pm7.0$  & N/A   \\ 
\hline
\end{tabular}
\caption{List of single top quark production cross section measurements at Tevatron and at the LHC.}
\label{tab:singletop}
\end{center}
\end{table}

The HL-LHC extrapolation of the observables presented in Tables~\ref{tab:whelicities},~\ref{tab:asymmetries} and~\ref{tab:singletop} was performed assuming a center-of-mass energy of 14~TeV and a total integrated luminosity of $3000$~fb$^{-1}$. The observable central values were set to their SM predictions. In the spirit of the recent ATLAS and CMS recommendations for the HL-LHC studies, the statistics driven sources and Monte Carlo event generation uncertainties were scaled according to the expected total integrated luminosity at the HL-LHC~\cite{Simone2018talk}. In practice the Monte Carlo related uncertainties become essentially negligible when compared with the other sources of error, given the foreseen large simulation sets. The theoretical and modelling uncertainties were tentatively extrapolated to half of the current value, once a better theoretical understanding of the studied processes is expected at HL-LHC. Intrinsic detector related uncertainties such as jets, electrons and photons energy scales and resolutions, muon and tau identification and reconstruction, or missing momentum and missing energy reconstruction, were considered to maintain their current value. While better reconstruction methods could be envisaged in the longer term, the fact that the events are detected in the busy environment of the HL-LHC is expected to degrade the overall reconstruction performance. The tagging of jets from the hadronization of $b$-quarks ($b-tagging$) is expected to be improved by a factor two, using novel techniques foreseen for the HL-LHC~\cite{Simone2018talk}. The extrapolated uncertainties, together with the observables central values, were included in the global fit in combination with the current measurements, in order to estimate the expected limits on the real and imaginary components of the top quark anomalous couplings. Since the future methods for object reconstruction at the HL-LHC are expected to be intrinsically different than the current ones, no correlation between the HL-LHC extrapolations and the current measurements were considered in this exercise.

\section{Results}
\label{sec:measurements}

The global fit to the top quark anomalous couplings presented in the $Wtb$ vertex Lagrangian~(\ref{ec:lagr}) was performed with {\sc TopFit}~\cite{topfit} for different scenarios, corresponding to different combinations of the $W$ boson helicity fractions, single top quark production cross sections and forward-backward asymmetries measurements. Limits on the anomalous couplings were set at 95\%~CL, assuming a top quark mass of $m_t = 175$~GeV, a bottom quark mass of $m_b=4.8$~GeV and a $W$ boson mass of $m_W = 80.4$~GeV, and allowing all real and imaginary components of these couplings to vary simultaneously, with the exception of $V_L = 1$. The impact of allowing $V_L$ to vary in the global fit has been shown to be small in recent studies, in particular for the most constrained coupling, \emph{i.e.}  $g_R$~\cite{Bernardo:2014vha,Birman:2016jhg,Deliot:2017byp}.

\begin{figure*}
\begin{center}
\vspace*{2cm}
\includegraphics[width=9cm]{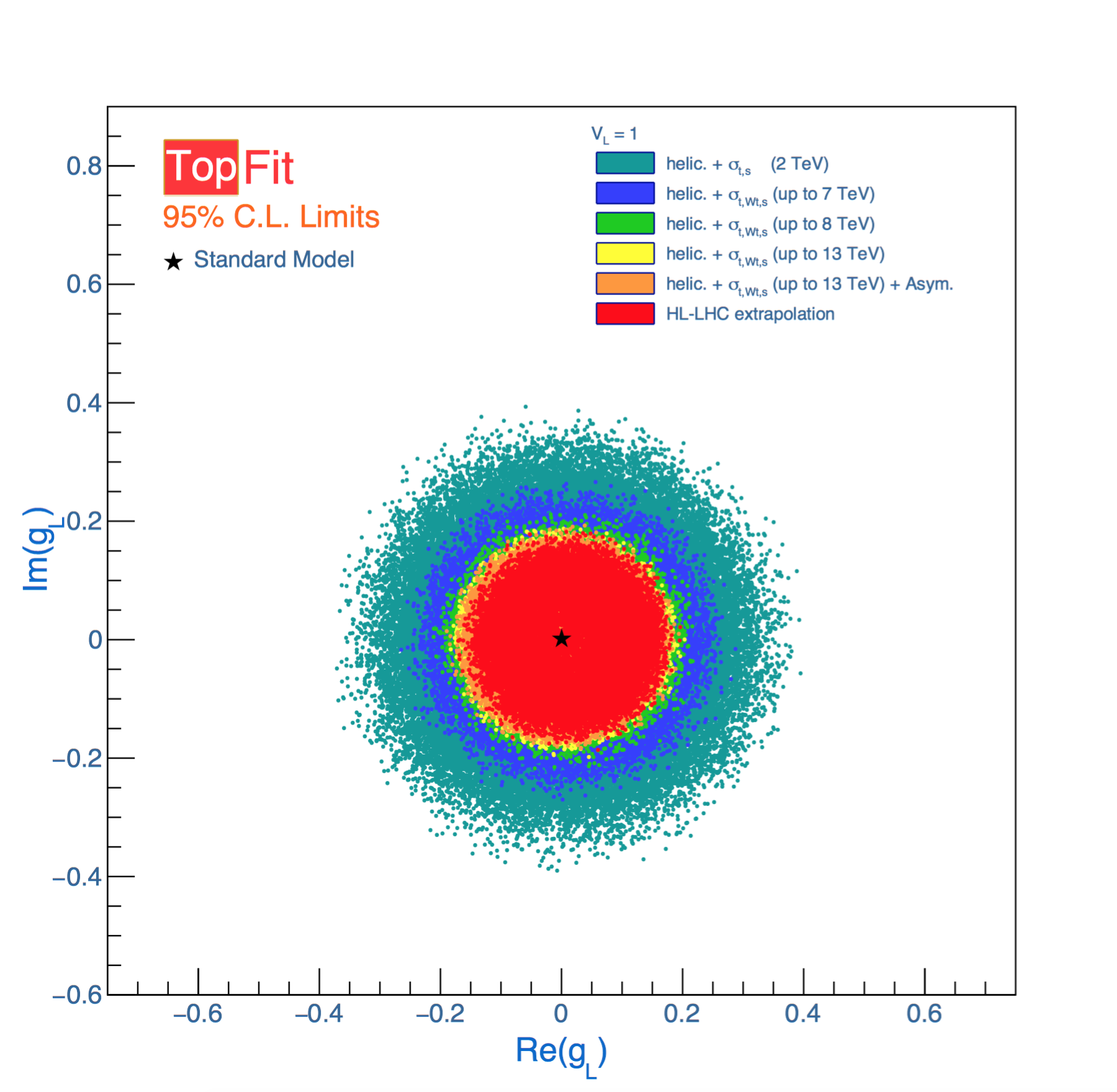}\includegraphics[width=9cm]{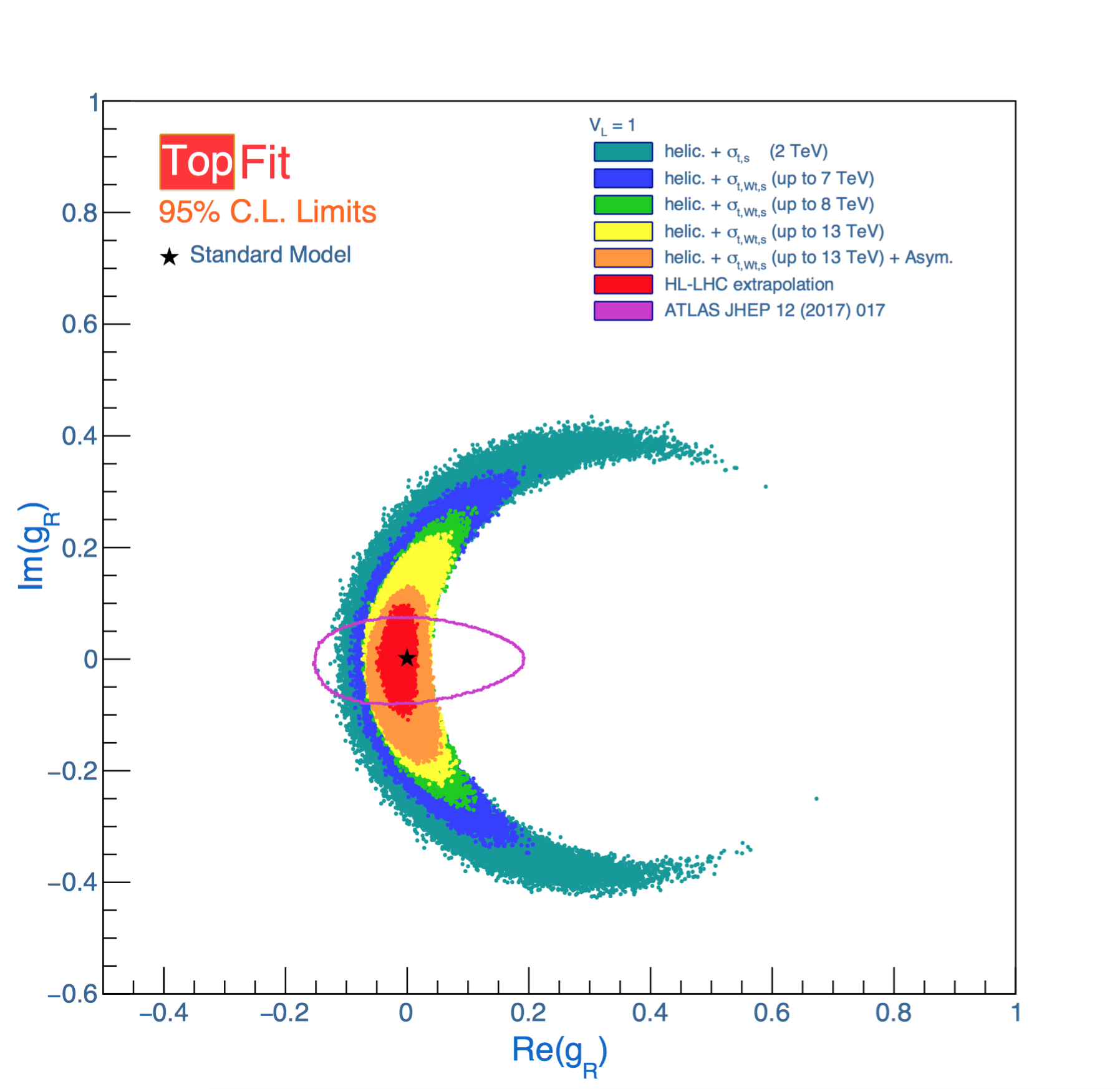} \\
\includegraphics[width=9cm]{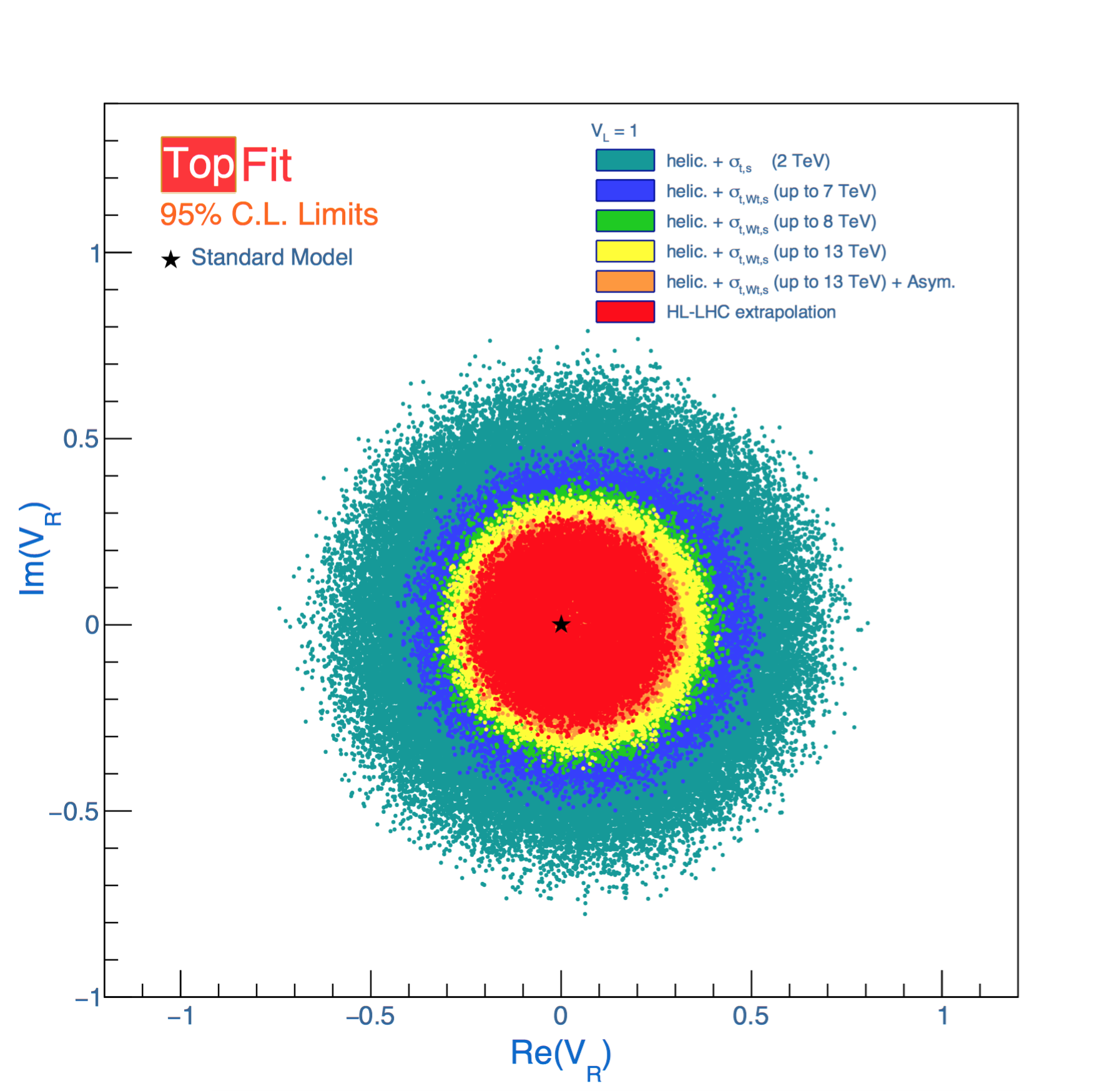}\includegraphics[width=9cm]{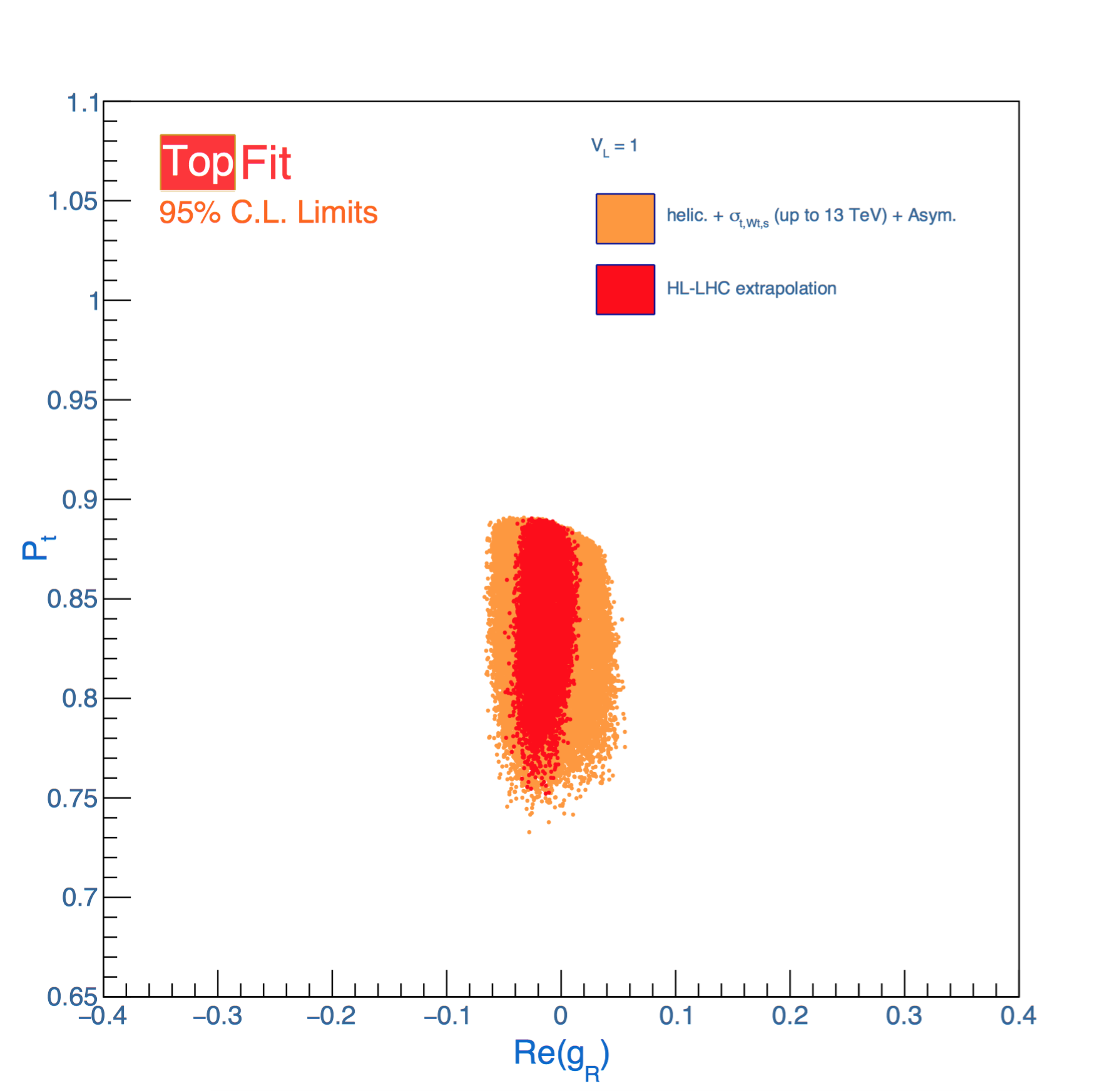} \\
\caption{Limits at 95\% CL on the allowed regions for the top quark anomalous couplings. Two-dimensional distributions are shown for the real versus the imaginary components of $g_L$ (upper left), $g_R$ (upper right) and $V_R$ (lower left) anomalous couplings. The top quark polarization is also shown as a function of the real part of $g_R$ (lower right). }
\label{fig:observables}
\end{center}
\end{figure*}

The allowed regions for the anomalous couplings are presented in Figure~\ref{fig:observables}. The different colors in the plots represent the allowed regions extracted from different combination of measurements used in {\sc TopFit}. For instance, allowed regions represented in blue-green were established using $W$ boson helicity fractions combined with single top quark production cross section measurements at Tevatron only. The navy blue, the green and yellow regions were extracted using the Tevatron measurements progressively combined with the LHC results at 7~TeV, 8~TeV and 13~TeV, respectively. The region in orange corresponds to the combination of all measurements up to a center-of-mass of 13~TeV with the forward-backward asymmetries, and the red region corresponds to the combination of these measurements with the HL-LHC extrapolated results. As expected, a general improvement is observed whenever additional measurements are included in the global fit. 

The upper left plot in Figure~\ref{fig:observables} shows the allowed region for the real and imaginary parts of $g_L$. The impact of the LHC measurements in the global fit is clearly visible, however, the combination with the HL-LHC extrapolated measurements does not provide any significant improvement for this anomalous coupling. A similar behavior is observed in the lower left plot, representing the allowed region of the real and imaginary parts of $V_R$. The upper right plot shows the allowed region for the real and imaginary parts of $g_R$. A clear improvement is observed with the inclusion of the HL-LHC extrapolated measurements for both $Re(g_R)$ and $Im(g_R)$. For completeness, the likelihood profile of the allowed region obtained from the triple angular decay rates from single top quark events alone is represented as a purple line~\cite{Aaboud:2017yqf}. The complementarity of all measurements is clearly visible and their impact on the sensitivity of the $g_R$ anomalous coupling is expected to be significant.
Last but not least, the allowed region for the top quark polarization versus the real part of $g_R$ is also represented in the lower right plot. A significant impact on the real part of $g_R$ is again observed with the inclusion of HL-LHC extrapolated measurements in the global fit, and no remarkable effect is observed on the top quark polarization. A summary of the one-dimensional limits is presented in Table~\ref{tab:ReCoup}, at 95\%~CL, for the real and imaginary parts of $g_R$, $g_L$ and $V_R$. The limits were extracted from the combination of $W$ boson helicity fractions, single top quark production cross sections, and the forward-backward asymmetries measured at the Tevatron, at the LHC and extrapolated to the HL-LHC. While $g_L$ and $V_R$ show no significant improvement when the HL-LHC extrapolated results are included in the global fit, the range of the allowed values (at 95\% CL) for both the real and imaginary parts of $g_R$, improve by roughly a factor of two with respect to current limits~\cite{Deliot:2017byp}, where $Re(g_R)= [-0.07,0.07]$ and $Im(g_R)=[-0.20,0.15]$.
\begin{table}[!h]
\begin{center}
\begin{tabular}{|c|c|c|c|}
\hline
   HL-LHC                   &             &                &           \\ 
        (3000~fb$^{-1}$)  &      $\gr$      &       $\gl$        &     $\Vr$ \\
\hline
     Allowed Region ($Re$)   	& [-0.05 , 0.02]  &    [-0.17 , 0.19]  & [-0.28 , 0.32] \\
     Allowed Region ($Im$)   	& [-0.11 , 0.10]  &    [-0.19 , 0.18]  & [-0.30 , 0.30] \\
\hline
\end{tabular}
\caption{95\%~CL limits on the allowed regions of the real and imaginary components of the anomalous couplings. The limits were extracted from the combination of $W$ boson helicity fractions, single top quark production cross sections and forward-backward asymmetries measured at Tevatron and at the LHC with the extrapolated results from the HL-LHC (3000~fb$^{-1}$).}
\label{tab:ReCoup}
\end{center}
\end{table}
\section{Conclusions}
\label{sec:conclusions}

Limits on the top quark anomalous couplings were presented at 95\%~CL for  different sets of observables, including the current most precise measurements on single top quark production cross sections, $W$ boson helicity fractions and forward-backward asymmetries at Tevatron and at the LHC. An extrapolation exercise of these measurements at the \mbox{HL-LHC} was performed to estimate the impact on the top quark anomalous couplings of a possible future 3000~fb$^{-1}$ run at a center-of-mass energy of 14~TeV. A clear improvement is observed for both the real and imaginary parts of the $g_R$ anomalous coupling when the HL-LHC extrapolated measurements are included in the global fit, corresponding to, at least, a factor 2 gain in sensitivity. No significant improvements are observed on the expected sensitivity to the remaining anomalous couplings, nor on the top quark polarization, given the current set of observables used. It should, however, be stressed that all forward-backward asymmetries are expected to play a crucial role in narrowing the limits on the imaginary part of $g_R$ and should be progressively included in a global fit.

\section*{Acknowledgments}

The authors would like to thank the Center for Theoretical Physics of the Physics Department at the New York City College of Technology, City University of New York, for providing computing power from their High-Performance Computing Cluster. This work was supported by the PSC-CUNY Award 61085-00 49 and by the contract  SFRH/BSAB/139747/2018  from Funda\c{c}\~ao para a Ci\^encia e Tecnologia (FCT).


\end{document}